\documentclass{webofc}
\usepackage[varg]{txfonts}   
\usepackage{color} 
\usepackage{xcolor} 
\usepackage{relsize} 
\usepackage{dsfont}
\usepackage{bbold}

\def\cmath{\color{black}}
\def\ctxt{\color{black}}

\newcommand{\barket}[1]{\left|#1\right\rangle} 
\newcommand{\brabar}[1]{\left\langle#1\right|} 
\newcommand{\braket}[1]{\left\langle#1\right\rangle} 
\def\bmat{\begin{pmatrix}}
\def\emat{\end{pmatrix}}
\def\C{\mathbb{C}} 
\def\N{\mathbb{N}} 
\def\Q{\mathbb{Q}} 
\def\R{\mathbb{R}} 
\def\Z{\mathbb{Z}}
\newcommand{\cabs}[1]{\left|#1\right|} 
\def\denmat{\mathlarger{\rho}}
\DeclareMathOperator{\e}{e}
\def\energyp{\boldsymbol{\varepsilon}}
\def\Entropy{\mathrm{\mathbf{S}}} 
\def\Hspace{\mathcal{H}} 
\DeclareMathOperator{\idmat}{\mathds{1}}

\newcommand{\inner}[2]{\left\langle#1\mid#2\right\rangle} 
\newcommand{\innerform}[3]{\braket{#1\cabs{#2}#3}} 
\DeclareMathOperator*{\KroneckerProduct}{\text{\raisebox{0.3ex}{$\mathsmaller{\bigotimes}$}}}
\def\lcycle{\boldsymbol{\ell}}	
\newcommand{\Math}[1]{$\cmath{}#1$} 
\newcommand{\MathEq}[1]{\begin{equation*}\cmath{#1}\end{equation*}}
\newcommand{\MathEqLab}[2]{\begin{equation}\cmath{#1}\label{#2}\end{equation}}

\newcommand{\MathEqArrLab}[1]{\begin{align}\cmath{}#1\end{align}}
\newcommand{\norm}[1]{\left\lVert#1\right\rVert}
\newcommand{\ordset}[1]{\left[#1\right]} 
\def\Prob{\mathrm{\mathbf{P}}} 
\newcommand{\projector}[1]{\Pi_{#1}}
\def\repq{\mathsf{U}} 
\def\runisymb{\mathsf{r}} 
\def\rcycle{\mathit{m}}	
\newcommand{\runi}[1]{\runisymb_{#1}} 
\newcommand{\set}[1]{\left\{#1\right\}} 
\DeclareMathOperator{\spn}{span}
\newcommand{\SymG}[1]{\mathsf{S}_{#1}} 
\newcommand{\timeder}[1]{\,\dot{#1}} 
\DeclareMathOperator{\tr}{tr}

\def\U{\mathsf{U}} 
\newcommand{\UG}[1]{\U\!\vect{#1}} 
\newcommand{\vect}[1]{\left(#1\right)} 
\newcommand{\Vthree}[3]{\bmat#1\\#2\\#3\emat} 
\def\wg{\mathsf{g}} 
\def\wG{\mathsf{G}} 
\def\wGN{\mathsf{M}} 
\def\wS{\Omega} 
\def\wSN{\mathsf{N}} 

\def\@oddhead{\hfil\thepage}
\def\mypar{\par\noindent$~~~~$}
\begin{document}
\pagestyle{myheadings}
\title{Modeling Quantum Behavior in the Framework of Permutation Groups}
%
%
%

\author{\vspace*{-10pt}\firstname{Vladimir} \lastname{Kornyak}\inst{1}\fnsep\thanks{\email{vkornyak@gmail.com}} 
}

\institute{Laboratory of Information Technologies,  Joint Institute for Nuclear Research, \\
141980 Dubna, Moscow Region, Russia
          }

\abstract{%
Quantum-mechanical concepts can be formulated in constructive finite terms without loss of their empirical content if we replace a general unitary group by a unitary representation of a finite group.
Any linear representation of a finite group can be realized as a subrepresentation of a permutation representation.
Thus, quantum-mechanical problems can be expressed in terms of permutation groups. 
This approach allows us to clarify the meaning of a number of physical concepts.
Combining methods of computational group theory with Monte Carlo simulation we study a model based on representations of permutation groups.
}
\vspace*{-15pt}
\maketitle
\def\ovsec{\vspace*{-9.6pt}}
\def\unsec{\vspace*{-4.6pt}}
\def\ovsubsec{\vspace*{-9.6pt}}
\def\unsubsec{\vspace*{-4.6pt}}
\vspace*{-5pt}
\ovsec
\section{Introduction}
\label{intro}
\unsec
Since the time of Newton, differential calculus demonstrates high efficiency in describing physical phenomena.
However, infinitesimal analysis introduces infinities in physical theories. This is often considered as a serious conceptual flaw:
recall, for example, Dirac's frequently quoted claim that the most important challenge in physics is ``to get rid of infinity''. 
Moreover, differential calculus, being, in fact, a kind of approximation, may lead to descriptive losses in some problems --- an illustrative example is given below in Sect.~\ref{subsecDIFF}.
In the paper, we describe a constructive version of quantum formalism that does not involve any concepts associated with actual infinities.
\par
The main part of the paper starts with Sect.~\ref{secQM}, which contains a summary of the basic concepts of the standard quantum mechanics with emphasis on the aspects important for our purposes.
\par
Sect.~\ref{secCoQM} describes a constructive modification of the quantum formalism.
We start with replacing a continuous group of symmetries of quantum states by a finite group.
The natural consequence of this replacement is unitarity, since any linear representation of a finite group is unitary.
Further, any finite group is naturally associated with some cyclotomic field. 
Generally, a cyclotomic field is a dense subfield of the field of complex numbers. 
This can be regarded as an explanation of the presence of complex numbers in the quantum formalism.
Any linear representation of a finite group over the associated cyclotomic field can be obtained from a permutation action of the group on vectors with natural components by projecting into suitable invariant subspace.
All this allows us to reproduce all the elements of quantum formalism in invariant subspaces of permutation representations.
\par
In Sect.~\ref{secModel} we consider a model of quantum evolution inspired by the quantum Zeno effect
--- the most convincing manifestation of the role of observation in the dynamics of quantum systems.
The model represents the quantum evolution as a sequence of observations with unitary transitions between them.
Standard quantum mechanics assumes a single deterministic unitary transition between observations.
In our model we generalize this assumption.
We treat a unitary transition as a kind of gauge connection --- a way of identifying indistinguishable entities at different times.
\emph{A priori}, any unitary transformation can be used as a data identification rule.
So, we assume that all unitary transformations participate in transitions between observations with appropriate weights.
We call a unitary evolution \emph{dominant} if it provides the maximum transition probability.%
\footnote{In fact, the principle of least action in physical theories implies the selection of dominant evolutions among all possible (``virtual'') evolutions.
The apparent determinism of these evolutions can be explained by the sharpness of their dominance.}
The Monte Carlo simulation shows a sharp dominance of such evolutions over other evolutions.
To compare with a continuous description, we present also the Lagrangian of the continuum approximation of the model.
\ovsec
\ovsec
\section{Formalism of quantum mechanics}\label{secQM}
\unsec
Here is a brief outline of the basic concepts of quantum mechanics.
We divide these concepts into three categories:
{\emph{states}}, {\emph{observations and measurements}}, and {\emph{time evolution}}.
\ovsubsec
\subsection{States}\label{subsecS}
\unsubsec
\mypar
A \emph{\textbf{pure quantum state}} is a ray in a Hilbert space \Math{\Hspace} over the complex field \Math{\C}, 
i.e. an \emph{equivalence class} of vectors \Math{\barket{\psi}\in\Hspace} with respect to the \emph{equivalence relation} \Math{\barket{\psi}\sim{}a\barket{\psi}}, where \Math{a\in\C,~a\neq0}.
We can reduce the equivalence classes by \emph{normalization}: \Math{\barket{\psi}\sim{}\e^{\mathrm{i}\alpha}\barket{\psi}, \norm{\psi}=1, \alpha\in\R}.
Finally, we can eliminate the phase ``degree of freedom'' \Math{\alpha} by transition to the rank one \emph{projector} \Math{\projector{\psi}=\barket{\psi}\!\brabar{\psi}},
which is a special case of a \emph{density matrix}.
\mypar
A \emph{\textbf{mixed quantum state}}  is described by a \emph{general density matrix} \Math{\denmat} characterized by the properties: 
(a)~\Math{\denmat=\denmat^\dagger},  (b)~\Math{\innerform{\psi}{\denmat}{\psi}\geq0} for any \Math{\barket{\psi}\in\Hspace},  (c)~\Math{\tr\denmat=1}.
In fact, any mixed state is a weighted mixture of pure states, i.e. its density matrix can be represented as a weighted sum of the rank one projectors.
We will denote the set of all density matrices by \Math{\mathcal{D}\!\vect{\Hspace}}.
\mypar
The Hilbert space of a \emph{\textbf{composite system}}, \Math{XY=X\times{}Y}, is the tensor product of the Hilbert spaces for the constituents: 
\Math{\Hspace_{{XY}}=\Hspace_{{X}}{\KroneckerProduct}\Hspace_{{Y}}}.
The \emph{\textbf{states of composite system}}, \Math{\mathcal{D}\!\vect{\Hspace_{{XY}}}}, are classified into two types: 
\emph{separable} and \emph{entangled} states.
The set of \emph{\textbf{separable}} states, \Math{\mathcal{D}_\mathrm{S}\!\vect{\Hspace_{{XY}}}}, consists of the states \Math{\denmat_{XY}\in\mathcal{D}\!\vect{\Hspace_{{XY}}}} 
that can be represented as weighted sums of the tensor products of states of the constituents:
\Math{\denmat_{{XY}}=\sum_{k}w_k\denmat_{{X}}^k{\KroneckerProduct}\denmat_{{Y}}^k,~~w_k\geq0,~~\sum_{k}w_k=1}.
The set of \emph{\textbf{entangled}} states, \Math{\mathcal{D}_\mathrm{E}\!\vect{\Hspace_{{XY}}}}, 
is by definition the complement of \Math{\mathcal{D}_\mathrm{S}\!\vect{\Hspace_{{XY}}}} in the set of all states:
\Math{\mathcal{D}_\mathrm{E}\!\vect{\Hspace_{{XY}}}=\mathcal{D}\!\vect{\Hspace_{{XY}}}\setminus\mathcal{D}_\mathrm{S}\!\vect{\Hspace_{{XY}}}}.
\ovsubsec
\subsection{Observations and measurements}\label{subsecOM}
\unsubsec
The terms `observation' and `measurement' are often used as synonyms.
However, it makes sense to separate these concepts: 
we treat observation as a more general concept which does not imply, in contrast to measurement, obtaining numerical information. 
\mypar
\emph{\textbf{Observation}} is the detection (``click of detector'') of a system, that is in the state \Math{\denmat}, in the subspace \Math{\mathcal{S}\leq\Hspace}.
The mathematical abstraction of the ``detector in the subspace'' \Math{\mathcal{S}} of a Hilbert space is the operator of projection, \Math{\projector{\mathcal{S}}}, into this subspace.
The result of quantum observation is random and its statistics is described by a probability measure defined on subspaces of the Hilbert space.
Any such measure \Math{\mu\vect{\cdot}} must be additive on any set of mutually orthogonal subspaces of a Hilbert space: if, e.g., \Math{\mathcal{A}} and \Math{\mathcal{B}}
are mutually orthogonal subspaces, then \Math{\mu\vect{\spn\vect{\mathcal{A}, \mathcal{B}}}=\mu\vect{\mathcal{A}}+\mu\vect{\mathcal{B}}}.
Gleason proved \cite{Gleason} that, excepting the case \Math{\dim\Hspace=2}, 
the only such measures have the form \Math{\mu_\denmat\vect{\mathcal{S}}=\tr\vect{\denmat\projector{\mathcal{S}}}}, where \Math{\denmat} is an arbitrary density matrix.
If, in particular, \Math{\denmat} describes a pure state, \Math{\denmat=\barket{\psi}\!\brabar{\psi}}, and \Math{\mathcal{S}} is one-dimensional, \Math{\mathcal{S}=\spn\vect{\barket{\varphi}}}, we come to the familiar Born rule: \Math{\tr\vect{\denmat\projector{\mathcal{S}}}=\Prob_{\text{Born}}=\cabs{\inner{\varphi}{\psi}}^2}.
\mypar
\emph{\textbf{Measurement}} is a special case of observation, 
when the partition of a Hilbert space into mutually orthogonal subspaces is provided by a Hermitian operator \Math{A}.
Any such operator can be written as \Math{A=\sum_ka_k\projector{e_k}}, where \Math{a_1, a_2,\ldots\in\R} is the \emph{spectrum} of \Math{A}, and \Math{e_1, e_2,\ldots} is an orthonormal basis of \emph{eigenvectors} of \Math{A}. 
``Click of the detector'' \Math{\projector{e_k}} is interpreted as that the \emph{eigenvalue} \Math{a_k} is the result of the measurement.
The mean for multiple measurements tends to the \emph{expectation value} of \Math{A} in the state \Math{\denmat}: 
\Math{\left\langle{A}\right\rangle_{\!\denmat}=\tr\vect{\denmat{A}}}.
\subsection{Time evolution}\label{subsecTE}
\unsubsec
\mypar
The \emph{\textbf{time evolution}} of a quantum system is a unitary transformation of data between observations. 
For a density matrix, unitary evolution takes the form
~\\[-10pt]
\MathEqLab{\denmat_{t'}=U_{t't}\denmat_tU_{t't}^{\dagger},}{EvoRho}
where \Math{\denmat_{t}} is the state \emph{after} observation at the time \Math{t},~
 \Math{\denmat_{t'}} is the state \emph{before} observation at the time \Math{t'}, and \Math{U_{t't}} is the unitary transition between the observation times \Math{t} and \Math{t'}.
In standard quantum formalism, time is considered as a continuous parameter, and relation \eqref{EvoRho} becomes the \emph{von Neumann equation} in the infinitesimal limit.
The evolution of a pure state can be written as \Math{\barket{\psi_{t'}}=U_{t't}\barket{\psi_t}}, and the corresponding infinitesimal limit is the \emph{Schr\"odinger equation}.
To emphasize the role of observation in quantum physics, we note that unitary evolution is simply a change of coordinates in Hilbert space and is not sufficient to describe observable physical phenomena. 
\ovsubsec
\subsection{Emergence of geometry within large Hilbert space via entanglement}\label{subsecEG}
\unsubsec
Quantum-mechanical theory does not need a geometric space as a fundamental concept --- everything can be formulated using only the Hilbert space formalism.
In this view, the observed geometry must emerge as an approximation.
The currently popular idea \cite{Raamsdonk,MalSus,Cao} of the emergence of geometry within a Hilbert space is based on the notion of entanglement.
Briefly, the scheme of extracting geometric manifold from the entanglement structure of a quantum state \Math{\denmat} in a Hilbert space \Math{\Hspace} is as follows:
\begin{itemize}
	\item
The Hilbert space decomposes into a large number of tensor factors:  \Math{\Hspace=\KroneckerProduct_x\Hspace_x,~ x\in{X}}.
Each factor is treated as a point (or bulk) of geometric space to be built.
A graph \Math{G} --- called \emph{tensor network} ---  with vertices \Math{x\in{X}} and edges \Math{\set{x,y}\in{}X\times{}X} is introduced.
	\item
The edges of \Math{G} are assigned \emph{weights} based on a \emph{measure of entanglement}, a function that vanishes on separable states and is positive on entangled states.
A typical such measure is the \emph{mutual information}:
\Math{\mathlarger{I}\!\vect{\denmat_{xy}}=S\!\vect{\denmat_{x}}+S\!\vect{\denmat_{y}}-S\!\vect{\denmat_{xy}}},~ where~\Math{\denmat_{x}} denotes the result of taking traces of \Math{\denmat} over all tensor factors excepting the \Math{x}-th (and similarly for \Math{\denmat_{x},} \Math{\denmat_{xy}});~ \Math{S\!\vect{a}=-\tr\vect{a\log{a}}} is the \emph{von Neumann entropy}.
The graph \Math{G} is supplied with a metric derived from the weights of the edges.
	\item
Finally, the graph \Math{G} is approximately isometrically embedded in a smooth metric manifold of as small as possible dimension using algorithms like \emph{multidimensional scaling} (MDS).
\end{itemize}
\ovsec
\section{Constructive modification of quantum formalism}\label{secCoQM}
\unsec
David Hilbert, a prominent advocate of the free use of the concept of infinity in mathematics,
wrote the following about the relation of the infinite to the reality:
``\emph{Our principal result is that the {infinite} is nowhere to be found in reality.
It {neither exists in nature} {nor provides} a legitimate {basis for rational thought} --- a remarkable harmony between being and thought.}''
Adopting this view, we reformulate the quantum formalism in constructive finite terms without distorting its empirical content \cite{Kornyak16,Kornyak15,Kornyak13}.
\ovsubsec
\subsection{Losses due to continuum and differential calculus}\label{subsecDIFF}
\unsubsec
Differential calculus (including differential equations, differential geometry, etc.) forms the basis of mathematical methods in physics.
The applicability of differential calculus is based on the assumption that any relevant function can be approximated by linear relations at small scales.
This assumption simplifies many problems in physics and mathematics, but at the cost of loss of completeness.
\par
As an example, consider the problem of classifying simple groups.
The concept of a group is an abstraction of the properties of \emph{permutations} (also called \emph{one-to-one mappings} or \emph{bijections}) of a set.
Namely, an abstract 
group is a set with an \emph{associative} operation, an \emph{identity} element, and an \emph{invertibility} for each element.
There are two most common additional assumptions that make the notion of a group  more meaningful: (a) \emph{the group is a differentiable manifold} --- such a group is called \emph{Lie group};
(b) \emph{the group is finite}.
It is clear that empirical physics is insensitive to assumption (b) --- ultimately, any empirical description is reduced to a finite set of data.
On the contrary, assumption (a) implies severe constraints on possible physical models.
\par
The problem of classification of simple groups%
\footnote{\emph{Simple groups}, i.e. groups that do not contain nontrivial \emph{normal subgroups}, are ``building blocks'' for all other groups.}
under assumption (a) turned out to be rather easy and was solved by two people (Killing and Cartan) in a few years.
The result is four infinite series: \Math{A_n}, \Math{B_n}, \Math{C_n}, \Math{D_n}; and five exceptional groups: \Math{E_6}, \Math{E_7}, \Math{E_8}, \Math{F_4}, \Math{G_2}.
\par
The solution of the classification problem under assumption (b) required the efforts of about a hundred people for over a hundred years \cite{Solomon}.
But the result --- \emph{``the enormous theorem''} --- turned out to be much richer.
The list of finite simple groups contains \Math{16+1+1} infinite series:
\begin{itemize}
\unsec
	\item \emph{groups of Lie type}: \\
\Math{A_n(q)}, \Math{B_n(q)}, \Math{C_n(q)}, \Math{D_n(q)}, \Math{E_6(q)}, \Math{E_7(q)}, \Math{E_8(q)}, \Math{F_4(q)}, \Math{G_2(q)},\\[2pt]
\Math{^2A_n\vect{q^2}}, \Math{^2B_n\vect{2^{2n+1}}}, \Math{^2D_n\vect{q^2}}, \Math{^3D_4\vect{q^3}},
\Math{^2E_6\vect{q^2}}, \Math{^2F_4\vect{2^{2n+1}}}, \Math{^2G_2\vect{3^{2n+1}}};
	\item \emph{cyclic groups of prime order},~ \Math{\Z_p};
	\unsec
	\item \emph{alternating groups},~ \Math{A_n,~n\geq5};
\end{itemize}
\unsec
and \Math{26} \emph{sporadic groups}:
\Math{M_{11}}, \Math{M_{12}}, \Math{M_{22}}, \Math{M_{23}}, \Math{M_{24}}, \Math{J_1}, \Math{J_2}, \Math{J_3}, \Math{J_4}, \Math{Co_1}, \Math{Co_2}, \Math{Co_3},\\
\phantom{and \Math{26} \emph{sporadic groups}~\,:}\Math{Fi_{22}}, \Math{Fi_{23}}, \Math{Fi_{24}}, \Math{HS}, \Math{McL}, \Math{He}, \Math{Ru}, \Math{Suz}, \Math{O'N}, \Math{HN},
\Math{Ly}, \Math{Th}, \Math{B}, \Math{M}.\\
Note that finite groups have an advantage over Lie groups in the sense that in empirical applications any Lie group can be modeled by some finite group, but not vice versa. 
\ovsubsec
\subsection{Replacing unitary group by finite group}\label{subsecRemInf}
\unsubsec
The main non-constructive element of the standard quantum formalism is the unitary group \Math{\UG{n}}, a set of cardinality of the continuum.
\par
Formally, the group \Math{\UG{n}} can be replaced by some finite group which is empirically equivalent to \Math{\UG{n}} as follows.
From the theory of quantum computing it is known that \Math{\UG{n}} contains a dense finitely generated --- and, hence, countable --- matrix subgroup \Math{\U_*\!\vect{n}}.
The group \Math{\U_*\!\vect{n}} is \emph{residually finite}, i.e. it has a reach set of non-trivial homomorphisms to finite groups.
\par
In essence, it is more natural to assume that at the fundamental level there are finite symmetry groups, 
and \Math{\UG{n}}'s are just continuum approximations of their unitary representations.
\par
The following properties of finite groups are important for our purposes:
\vspace*{-4pt}
\begin{itemize}
	\item
any finite group is a subgroup of a \emph{symmetric group},
\item
\vspace*{-4pt}
any linear representation of a finite group is \emph{unitary},
\item
\vspace*{-4pt}
any linear representation is  subrepresentation of some \emph{permutation representation}.
\end{itemize}
\ovsubsec
\subsection{``Physical'' numbers}\label{subsecPhysNumb}
\unsubsec
The basic number system in quantum formalism is the complex field \Math{\C}.
This non-constructive field can be obtained as a metric completion of many algebraic extensions of rational numbers.
We consider here constructive numbers that are closely related to finite groups and are based on two primitives with a clear intuitive meaning:
\begin{enumerate}
	\item
\vspace*{-6pt}	
\emph{natural numbers} (``counters''): \Math{\N=\set{0,1,\ldots}};
	\item
\vspace*{-6pt}	
\Math{k}th \emph{roots of unity}%
\footnote{There are \Math{k} different \Math{k}th roots of unity.
A \Math{k}th root of unity is called \emph{primitive} if \Math{\runi{k}^m\neq1} for any \Math{m:~0<m<k}.}
 (``algebraic form of the idea of \Math{k}-periodicity''): \Math{\runi{k}\mid\runi{k}^k=1}.
\end{enumerate}
\vspace*{-6pt}
These basic concepts are sufficient to represent all physically meaningful numbers.
\par
We start by introducing \Math{\N\!\ordset{\runi{k}}}, the extension of the \emph{semiring} \Math{\N} by primitive \Math{k}th root of unity.
\Math{\N\!\ordset{\runi{k}}} is a \emph{ring} if \Math{k\geq2}.
This construction allows, in particular, to add \emph{negative numbers} to the naturals: 
\Math{\Z=\N\!\ordset{\runi{2}}} is the extension of \Math{\N} by the primitive square root of unity.
Further, by a standard mathematical procedure, we obtain the \Math{k}th \emph{cyclotomic field} \Math{\Q\!\vect{\runi{k}}} as the fraction field of the ring \Math{\N\!\ordset{\runi{k}}}.
If \Math{k\geq3}, then the field \Math{\Q\!\vect{\runi{k}}} is a \emph{dense subfield} of \Math{\C}, i.e. (constructive) cyclotomic fields are empirically indistinguishable from the (non-constructive) complex field. Note that \Math{\Q\cong\Q\!\vect{\runi{2}}}.
\par
The importance of cyclotomic numbers for constructive quantum mechanics is explained by the following.
Let us recall some terms.
The \emph{exponent} of a group \Math{\wG} is the least common multiple of the orders of its elements.
A \emph{splitting field} for a group \Math{\wG} is a field that allows to split completely any linear representation of \Math{\wG} into irreducible components.
A \emph{minimal splitting field} is a splitting field that does not contain proper splitting subfields.
Although minimal splitting field for a given group \Math{\wG} may be non-unique, any minimal splitting field is a subfield of some cyclotomic field \Math{\Q\!\vect{\runi{k}}}, 
where \Math{k} is a divisor of the exponent of \Math{\wG}. 
Thus, to work with any unitary representation of \Math{\wG} it is sufficient to use the \Math{k}th cyclotomic field, where \Math{k} is related to the structure of \Math{\wG}.
\ovsubsec
\subsection{Constructive representations of a finite group}\label{subsecConstrRep}
\unsubsec
Let a group \Math{\wG} act by permutations on a set \Math{\wS,~ \cabs{\wS}=\wSN}.
If we assume that the elements of \Math{\wS} are ``types'' of some discrete entities ({``ontological entities'', ``elements of reality''}), 
then the collections of these entities can be described as elements of the module  \Math{H=\N^\wSN} over the semiring \Math{\N} with the basis \Math{\wS}.
The decomposition of the action of \Math{\wG} in the module \Math{H} into irreducible components reflects the structure of the invariants of the action.
In order for the decomposition to be complete, it is necessary to extend the semiring \Math{\N} to a splitting field, 
e.g., to a cyclotomic field \Math{\Q\vect{\runi{k}}}, where \Math{k} is a suitable divisor of the exponent of \Math{\wG}. 
With such an extension of the scalars, the module \Math{H} is transformed into the Hilbert space \Math{\Hspace} over \Math{\Q\vect{\runi{k}}}.
This construction, with a suitable choice of the permutation domain \Math{\wS}, 
allows us to obtain \emph{any representation} of the group \Math{\wG} in some invariant subspace of the Hilbert space \Math{\Hspace}.
We obtain ``quantum mechanics'' within an invariant subspace if, in addition to unitary evolutions, projective measurements are also restricted by this subspace.
\par
\begin{figure}[h]
\centering
\sidecaption
\includegraphics[width=0.55\textwidth]{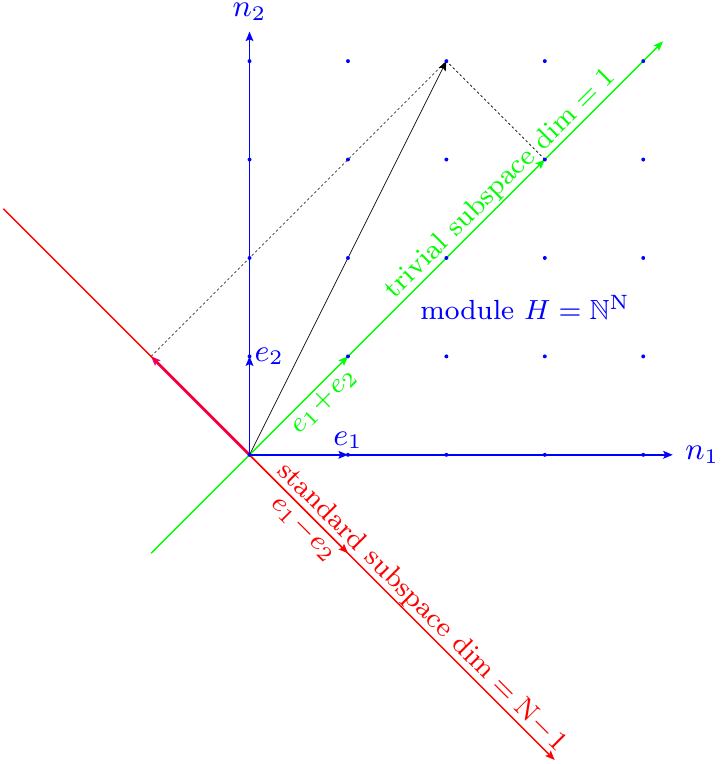}
\caption{Natural representation of \Math{\SymG{{\wSN}}} decomposes into two irreducibles: \Math{1}D \emph{trivial} and \Math{\vect{\wSN-1}}D \emph{standard} \vspace*{2pt}representations.
Canonical bases:\hspace{105pt}
in trivial subspace\hspace{100pt}\Math{~~~e_1+e_2+\cdots+e_{\wSN}}
\hspace{105pt}in standard subspace
\hspace{105pt}\Math{~~~e_1-e_2}
\hspace{130pt}\Math{~~~e_2-e_3}
\hspace{139pt}\Math{~~~~~~~~\vdots}
\hspace{130pt}\Math{~~~e_{\wSN-1}-e_{\wSN}}
}
\label{NatStdS2}       
\end{figure}

The above is illustrated in Figure \ref{NatStdS2} by the example of the natural action of the symmetric group \Math{\SymG{{\wSN}}} on the set
\Math{\wS=\set{e_1,\ldots,e_{\wSN}}}. 
Note, that any symmetric group is a \emph{rational-representation} group, 
i.e. the field of rational numbers \Math{\Q} is a splitting field for \Math{\SymG{{\wSN}}}. 
%
\section{Modeling quantum evolution}\label{secModel}
The fundamental discrete time \Math{\mathcal{T}} is represented by an ordered sequence of 
integers:
\Math{\mathcal{T}=\N} or
\Math{\mathcal{T}=\Z}.
We define a finite sequence of ``instants of observations''  as a subsequence of \Math{\mathcal{T}}:
\MathEqLab{\ordset{t_{0},t_{1},\ldots,t_{k-1},t_{k},\ldots,t_n}.}{Tobserv}
The data of the model of quantum evolution include the sequence of the length \Math{n+1} for states
\MathEqLab{\ordset{\denmat_{0},\denmat_{1},\ldots,\denmat_{k-1},\denmat_{k},\ldots,\denmat_n}}{seqStates}
and the sequence of the length \Math{n} for unitary transitions between observations
\MathEqLab{\ordset{U_{1},\ldots,U_{k},\ldots,U_n}.}{seqU}
Standard quantum mechanics presupposes a single unitary evolution, \Math{U_k}, between observations at times \Math{t_{k-1}} and \Math{t_k}.
The \emph{single-step transition probability} takes the form
\MathEqLab{\Prob_k=\tr\vect{U_k\denmat_{k-1}U_k^\dagger\denmat_k}.}{ProbStd}
The evolution can be expressed via the Hamiltonian: \Math{U_{k}=\e^{-\mathrm{i}{H}\vect{t_k-t_{k-1}}}}.
In physical theories, Hamiltonians are usually derived from \emph{the principle of least action},
which, like any extremal principle, implies the selection of a small subset of dominant elements in a large set of candidates.
Thus it is natural to assume that, in fact,  all unitary evolutions take part in the transition between observations with their weights, but only the dominant evolutions are manifested in observations.
Therefore, in our model, we use the following modification of the single-step transition probability
\MathEqLab{\Prob_k=\sum_{m=1}^{\wGN}w_{km}\tr\vect{U_{k,m}\denmat_{k-1}U_{k,m}^\dagger\denmat_k},}{ProbMod}
where \Math{U_{k,m}=\repq\vect{g_m}},~ \Math{g_m\in\wG};~ 
\Math{\wG=\set{\wg_1,\ldots,\wg_\wGN}} is a finite group;  
\Math{\repq} is a  unitary representation of \Math{\wG};   
\Math{w_{km}} is the weight of \Math{m}th group element at \Math{k}th transition.
\par
The operators \Math{U_{k,m}}, that maximize 
\Math{\tr\vect{U_{k,m}\denmat_{k-1}U_{k,m}^\dagger\denmat_k}}, will be called \emph{dominant evolutions}.
~\\[-8pt]
\MathEqLab{\text{\ctxt{}The \emph{single-step entropy} is defined as}\hspace*{10pt}\Delta\Entropy_{k}\!=-\log\Prob_{k}.\hspace*{145pt}}{Entropy1}
~\\[-12pt]
Continuum approximation of \eqref{Entropy1} leads to the \emph{Lagrangian} \Math{\mathcal{L}}.
Taking the logarithm of the probability of the whole trajectory, \Math{\Prob_{0\rightarrow{n}}=\prod_{k=1}^n\Prob_k},
we arrive at the \emph{entropy of trajectory} \Math{\Entropy_{{0}\rightarrow{n}}\!=\sum_{k=1}^n\Delta\Entropy_{k}},
the continuum approximation of which is the \emph{action} \Math{\mathcal{S}=\int\!\mathcal{L}dt}. 
\ovsubsec
\subsection{Continuum approximation of discrete model}\label{subsecDiscrCont}
\unsubsec
Continuum approximation of the above model requires the following simplifying assumptions:
\begin{itemize}
	\item 
Sequence \eqref{Tobserv} should be replaced by a continuous time interval \Math{\ordset{t_0,t_n}\subseteq\R}.
	\item 
Sequences \eqref{seqStates} and \eqref{seqU} are to be replaced by continuous functions of time, \Math{\denmat=\denmat\vect{t}} and \Math{U=U\vect{t}}.
	\item 
The relation \Math{\tr\vect{\denmat^2}=1} is necessary to ensure the continuity of probability.
This relation holds only for pure states \Math{\denmat=\barket{\psi}\!\brabar{\psi}}. 
So, we will consider \Math{{\psi}} instead of \Math{\denmat}.
	\item 
Assuming that \Math{U} belongs to a unitary representation of a Lie group, we use the Lie algebra\\ approximation, \Math{U\approx\idmat+\mathrm{i}{}A}, where \Math{A=A\vect{t}} is a function whose values are Hermitian matrices.
	\item 
We introduce derivatives and use the linear approximations \Math{\Delta{A}\approx\timeder{A}\Delta{t}} and \Math{\Delta{\psi}\approx\timeder{\psi}\Delta{t}}.
\end{itemize}
Applying these assumptions and approximations to the single-step entropy \eqref{Entropy1} and taking the infinitesimal limit we obtain the Lagrangian:
\MathEq{\mathcal{L}=\underbrace{\innerform{\psi}{\timeder{A}^2}{\psi}-\innerform{\psi}{\timeder{A}}{\psi}^2}_{\text{\ctxt{}dispersion of}~\timeder{A}~\text{\ctxt{}in state}~\psi}-\mathrm{i}\bigg(\!\innerform{\timeder{\psi}}{\timeder{A}}{\psi}-\innerform{\psi}{\timeder{A}}{\timeder{\psi}}+2\innerform{\psi}{\timeder{A}}{\psi}\inner{\psi}{\timeder{\psi}}\!\bigg)-\inner{\psi\!}{\!\timeder{\psi}}^2.}
\ovsubsec
\subsection{Dominant unitary evolutions in symmetric group}\label{subsecDomEvol}
\unsubsec
The dominant evolutions between the states represented by the vectors from the module \Math{H=\N^\wSN} for the group \Math{\SymG{\wSN}} can be computed as follows.\\
Let \Math{\barket{n}=\Vthree{n_1}{\vdots}{n_\wSN},~\barket{m}=\Vthree{m_1}{\vdots}{m_\wSN},~\barket{1}=\Vthree{1}{\vdots}{1}}~~ be \Math{\wSN}-dimensional vectors with natural components.\\
The Born probabilities for the pair \Math{\barket{n}} and \Math{\barket{m}} are
\MathEqArrLab{\Prob_\mathrm{nat}\vect{\barket{n}\!,\barket{m}}=&\cmath\frac{\inner{n}{m}^2}{\inner{n}{n}\inner{m}{m}}~~~\text{\small\ctxt{}---~natural representation,}\nonumber\\
\cmath\Prob_\mathrm{std}\vect{\barket{n}\!,\barket{m}}=&\cmath\frac{\vect{\inner{n}{m}-\frac{1}{\wSN}\inner{n}{\!1}\inner{1\!}{m}}^2}
{\vect{\inner{n}{n}-\frac{1}{\wSN}\inner{n}{\!1}\inner{1\!}{n}}\vect{\inner{m}{m}-\frac{1}{\wSN}\inner{m}{\!1}\inner{1\!}{m}}}~~~\text{\small\ctxt{}---~standard representation.}\label{Pstd}}
Let \Math{R_a} denote the permutation (as well as its representation), that sorts the components of vector \Math{\barket{a}} in some order.
It is not hard to show that the unitary operator \Math{U=R_m^{-1}R_n} maximizes the probability \Math{\Prob_\mathrm{\!*}\!\vect{U\!\barket{n}\!,\barket{m}}},
where the permutations \Math{R_n} and \Math{R_m} sort the vectors  \Math{\barket{n}} and \Math{\barket{m}} \emph{identically} in the case of natural representation,
and either \emph{identically} or \emph{oppositely} --- depending on the value of the numerator in \eqref{Pstd} --- in the case of standard representation.
\ovsubsec
\subsection{Energy of permutation}\label{subsecPermEn}
\unsubsec
Planck's formula, \Math{E={h}\boldsymbol{\nu}}, relates energy to frequency.
This relation is reproduced by the quantum-mechanical definition of energy as an eigenvalue of the Hamiltonian, \Math{H={\mathrm{i}}{\hbar}\ln{}U}, associated with a unitary transformation.
Consider the energy spectrum of a unitary operator defined by a permutation.
Let \Math{p} be a permutation of the cycle type 
\Math{\set{{\lcycle_1^{\rcycle_1}},\ldots,{\lcycle_k^{\rcycle_k}},\ldots,{\lcycle_K^{\rcycle_K}}}}, 
where \Math{\lcycle_k} and \Math{\rcycle_k} represent lengths and multiplicities of cycles in the decomposition of \Math{p} into disjoint cycles.
A short calculation shows that the Hamiltonian of the permutation \Math{p} has the following diagonal form
 \MathEq{{}H_p=
\bmat
\idmat_{\rcycle_1}\!\otimes\,H_{\lcycle_1}
&&\\
&\hspace{-25pt}
\ddots&\\
&&
\hspace{-25pt}
\idmat_{\rcycle_K}\!\otimes\,H_{\lcycle_K}
\emat,~~~\text{\ctxt{}where}~~~H_{\lcycle_k}={\displaystyle\frac{1}{\lcycle_k}}
\bmat
0&&&\\[-4pt]
&\hspace{-6pt}{1}&&\\[-4pt]
&&\hspace{-6pt}\ddots&\\[-4pt]
&&&\hspace{-8pt}\lcycle_k\!-\!1
\emat.
}
We shall call the least nonzero energy of a permutation the \emph{base energy}:
\MathEqLab{\displaystyle\energyp
=\frac{1}{\max\vect{\lcycle_1,\ldots,\lcycle_K}}\,.}{Ebase}
Simulation 
shows that the base \emph{(``ground state'', ``zero-point'', ``vacuum'')} energy is statistically more significant than other energy levels.
\def\hsp{\hspace*{-40pt}}
\def\wid{1.1\textwidth}
\def\hei{0.165\textheight}
\begin{figure}[ht]
\centering
\hsp\includegraphics[width=\wid,height=\hei]{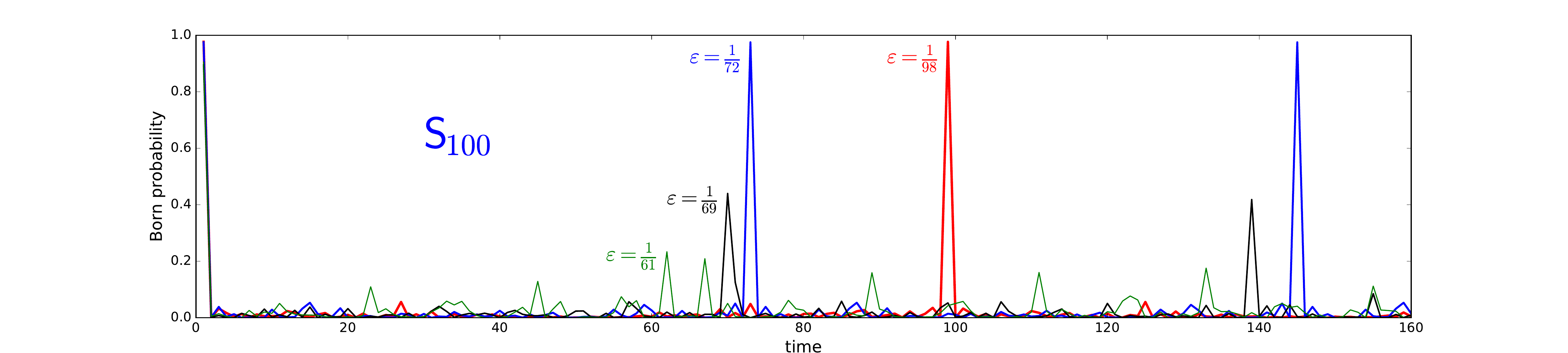}
\hsp\includegraphics[width=\wid,height=\hei]{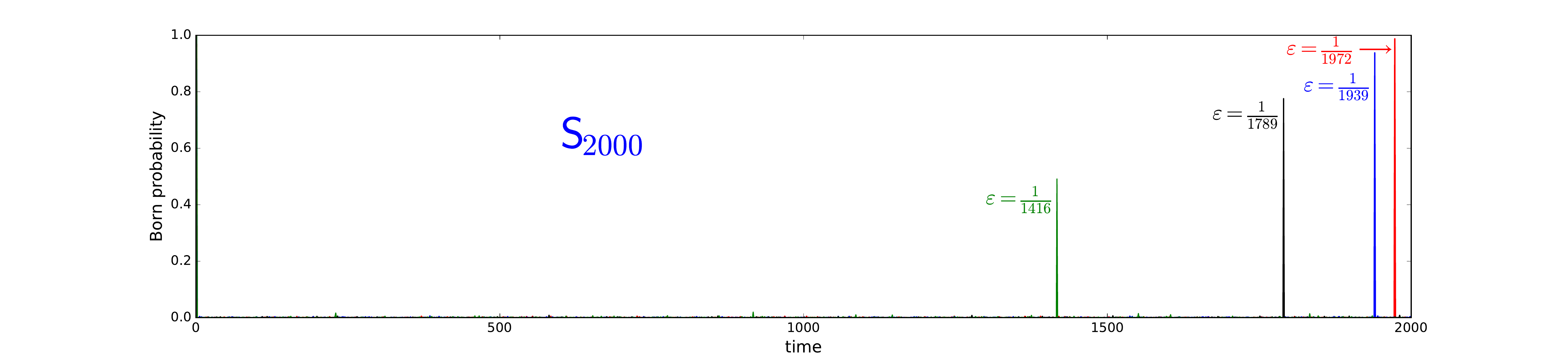}
\caption{Dominant evolutions between randomly generated states. Born probability vs time}
\label{MonteCarlo}       
\end{figure}
\ovsubsec
\subsection{Monte Carlo simulation of dominant evolutions}\label{subsecMonteCarlo}
\unsubsec
Figure \ref{MonteCarlo} shows several dominant evolutions for the standard representation of the groups 
\Math{\SymG{100}} and  \Math{\SymG{2000}}. 
Each graph represents the time dependencies of Born's probabilities for the dominant evolutions between four randomly generated pairs of natural vectors. 
The dominant evolutions are marked by labeling their peaks with their base energies: 
\Math{\energyp\in\set{\frac{1}{61}, \frac{1}{69}, \frac{1}{72}, \frac{1}{98}}} and \Math{\energyp\in\set{\frac{1}{1416}, \frac{1}{1789}, \frac{1}{1939}, \frac{1}{1972}}}
for \Math{\SymG{100}} and  \Math{\SymG{2000}}, respectively.
We see that with increasing the group size, non-dominant evolutions become almost invisible against the sharp peaks of dominant evolutions.
\ovsec
\section{Summary}\label{Summary}
\unsec
\begin{enumerate}
	\item 
A constructive version of quantum formalism can be formulated in terms of projections of permutations of finite sets into invariant subspaces.
	\item 
	\unsec
Quantum randomness is a consequence of the fundamental impossibility of tracing the individuality of indistinguishable entities in their evolution.	
	\item 
	\unsec
The natural number systems for quantum formalism are cyclotomic fields, and the field of complex numbers is just their non-constructive metric completion.	
	\item 
	\unsec
Observable behavior of quantum system is determined by the dominants among all possible quantum evolutions.
	\item 
	\unsec
The principle of least action is a continuum approximation of the principle of selection of the most probable trajectories.	
\end{enumerate}

\end{document}